\documentclass[preprint]{aastex631}

\shorttitle{Solar-Cycle Variations of SAA Protons Measured With PAMELA}
\shortauthors{Bruno et al.}

\begin{document}

\title{Solar-Cycle Variations of South-Atlantic Anomaly Proton Intensities Measured With The PAMELA Mission}

\correspondingauthor{Alessandro Bruno}
\email{alessandro.bruno-1@nasa.gov}

\author{A. Bruno}
\affiliation{Heliophysics Division, NASA Goddard Space Flight Center, Greenbelt, MD 20771, USA}
\affiliation{Department of Physics, Catholic University of America, Washington, DC 20064, USA}

\author{M. Martucci}
\affiliation{University of Rome ``Tor Vergata'', Department of Physics, I-00133 Rome, Italy}
\affiliation{INFN, Sezione di Rome ``Tor Vergata'', I-00133 Rome, Italy}

\author{F. S. Cafagna}
\affiliation{INFN, Sezione di Bari, I-70126 Bari, Italy}

\author{R. Sparvoli}
\affiliation{University of Rome ``Tor Vergata'', Department of Physics, I-00133 Rome, Italy}
\affiliation{INFN, Sezione di Rome ``Tor Vergata'', I-00133 Rome, Italy}


\author{O. Adriani} 
\affiliation{University of Florence, Department of Physics, I-50019 Sesto Fiorentino, Florence, Italy}
\affiliation{INFN, Sezione di Florence, I-50019 Sesto Fiorentino, Florence, Italy} 

\author{G. C. Barbarino} 
\affiliation{University of Naples ``Federico II'', Department of Physics, I-80126 Naples, Italy}  
\affiliation{INFN, Sezione di Naples, I-80126 Naples, Italy} 
 
\author{G. A. Bazilevskaya} 
\affiliation{Lebedev Physical Institute, RU-119991, Moscow, Russia} 
 
\author{R. Bellotti}
\affiliation{INFN, Sezione di Bari, I-70126 Bari, Italy}
\affiliation{University of Bari, Department of Physics, I-70126 Bari, Italy}

\author{M. Boezio}
\affiliation{INFN, Sezione di Trieste, I-34149 Trieste, Italy} 
\affiliation{IFPU, I-34014 Trieste, Italy} 

\author{E. A. Bogomolov}
\affiliation{Ioffe Physical Technical Institute, RU-194021 St. Petersburg, Russia} 

\author{M. Bongi} 
\affiliation{University of Florence, Department of Physics, I-50019 Sesto Fiorentino, Florence, Italy}
\affiliation{INFN, Sezione di Florence, I-50019 Sesto Fiorentino, Florence, Italy} 

\author{V. Bonvicini} 
\affiliation{INFN, Sezione di Trieste, I-34149 Trieste, Italy}

\author{D. Campana} 
\affiliation{INFN, Sezione di Naples, I-80126 Naples, Italy} 

\author{P. Carlson} 
\affiliation{KTH, Department of Physics, and the Oskar Klein Centre for Cosmoparticle Physics, AlbaNova University Centre, SE-10691 Stockholm, Sweden}

\author{M. Casolino}
\affiliation{INFN, Sezione di Rome ``Tor Vergata'', I-00133 Rome, Italy}  
\affiliation{RIKEN, EUSO team Global Research Cluster, Wako-shi, Saitama, Japan} 

\author{G. Castellini} 
\affiliation{IFAC, I-50019 Sesto Fiorentino, Florence, Italy}

\author{C. De Santis}
\affiliation{INFN, Sezione di Rome ``Tor Vergata'', I-00133 Rome, Italy}  

\author{A. M. Galper}
\affiliation{MEPhI: National Research Nuclear University MEPhI, RU-115409, Moscow, Russia} 

\author{S. V. Koldashov}\altaffiliation{Deceased} 
\affiliation{MEPhI: National Research Nuclear University MEPhI, RU-115409, Moscow, Russia}

\author{S. Koldobskiy} 
\affiliation{MEPhI: National Research Nuclear University MEPhI, RU-115409, Moscow, Russia} 
\affiliation{University of Oulu, 90570 Oulu, Finland}

\author{A. N. Kvashnin}
\affiliation{Lebedev Physical Institute, RU-119991, Moscow, Russia}

\author{A. Lenni}
\affiliation{University of Trieste, Department of Physics, I-34100 Trieste, Italy}
\affiliation{INFN, Sezione di Trieste, I-34149 Trieste, Italy} 
\affiliation{IFPU, I-34014 Trieste, Italy} 

\author{A.A. Leonov} 
\affiliation{MEPhI: National Research Nuclear University MEPhI, RU-115409, Moscow, Russia} 

\author{V.V. Malakhov} 
\affiliation{MEPhI: National Research Nuclear University MEPhI, RU-115409, Moscow, Russia} 

\author{L. Marcelli} 
\affiliation{INFN, Sezione di Rome ``Tor Vergata'', I-00133 Rome, Italy}  

\author{N. Marcelli} 
\affiliation{University of Rome ``Tor Vergata'', Department of Physics, I-00133 Rome, Italy}
\affiliation{INFN, Sezione di Rome ``Tor Vergata'', I-00133 Rome, Italy} 

\author{A. G. Mayorov} 
\affiliation{MEPhI: National Research Nuclear University MEPhI, RU-115409, Moscow, Russia} 

\author{W. Menn}
\affiliation{Universitat Siegen, Department of Physics, D-57068 Siegen, Germany}

\author{M. Merg\`e} 
\affiliation{INFN, Sezione di Rome ``Tor Vergata'', I-00133 Rome, Italy}  
\affiliation{University of Rome ``Tor Vergata'', Department of Physics, I-00133 Rome, Italy}

\author{E. Mocchiutti} 
\affiliation{INFN, Sezione di Trieste, I-34149 Trieste, Italy} 

\author{A. Monaco} 
\affiliation{INFN, Sezione di Bari, I-70126 Bari, Italy}
\affiliation{University of Bari, Department of Physics, I-70126 Bari, Italy}

\author{N. Mori} 
\affiliation{INFN, Sezione di Florence, I-50019 Sesto Fiorentino, Florence, Italy} 

\author{V. V. Mikhailov} 
\affiliation{MEPhI: National Research Nuclear University MEPhI, RU-115409, Moscow, Russia} 

\author{R. Munini} 
\affiliation{INFN, Sezione di Trieste, I-34149 Trieste, Italy} 
\affiliation{IFPU, I-34014 Trieste, Italy} 

\author{G. Osteria}
\affiliation{INFN, Sezione di Naples, I-80126 Naples, Italy} 

\author{B. Panico} 
\affiliation{INFN, Sezione di Naples, I-80126 Naples, Italy} 
 
\author{P. Papini} 
\affiliation{INFN, Sezione di Florence, I-50019 Sesto Fiorentino, Florence, Italy} 

\author{M. Pearce}
\affiliation{KTH, Department of Physics, and the Oskar Klein Centre for Cosmoparticle Physics, AlbaNova University Centre, SE-10691 Stockholm, Sweden}

\author{P. Picozza} 
\affiliation{University of Rome ``Tor Vergata'', Department of Physics, I-00133 Rome, Italy}
\affiliation{INFN, Sezione di Rome ``Tor Vergata'', I-00133 Rome, Italy}  

\author{M. Ricci}
\affiliation{INFN, Laboratori Nazionali di Frascati, Via Enrico Fermi 40, I-00044 Frascati, Italy}

\author{S. B. Ricciarini}
\affiliation{INFN, Sezione di Florence, I-50019 Sesto Fiorentino, Florence, Italy} 
\affiliation{IFAC, I-50019 Sesto Fiorentino, Florence, Italy}

\author{M. Simon}\altaffiliation{Deceased}
\affiliation{Universitat Siegen, Department of Physics, D-57068 Siegen, Germany}

\author{A. Sotgiu}
\affiliation{University of Rome ``Tor Vergata'', Department of Physics, I-00133 Rome, Italy}
\affiliation{INFN, Sezione di Rome ``Tor Vergata'', I-00133 Rome, Italy}

\author{P. Spillantini}
\affiliation{MEPhI: National Research Nuclear University MEPhI, RU-115409, Moscow, Russia} 
\affiliation{Istituto Nazionale di Astrofisica, Fosso del cavaliere 100, I-00133 Rome, Italy} 

\author{Y. I. Stozhkov} 
 \affiliation{Lebedev Physical Institute, RU-119991, Moscow, Russia} 

\author{A. Vacchi}
\affiliation{INFN, Sezione di Trieste, I-34149 Trieste, Italy} 
\affiliation{University of Udine, Department of Mathematics, Computer Science and Physics, Via delle Scienze, 206, Udine, Italy}

\author{E. Vannuccini}
\affiliation{INFN, Sezione di Florence, I-50019 Sesto Fiorentino, Florence, Italy} 

\author{G.I. Vasilyev} 
\affiliation{Ioffe Physical Technical Institute, RU-194021 St. Petersburg, Russia} 

\author{S. A. Voronov} 
\affiliation{MEPhI: National Research Nuclear University MEPhI, RU-115409, Moscow, Russia} 

\author{Y. T. Yurkin} 
\affiliation{MEPhI: National Research Nuclear University MEPhI, RU-115409, Moscow, Russia} 

\author{G. Zampa}
\affiliation{INFN, Sezione di Trieste, I-34149 Trieste, Italy} 

\author{N. Zampa}
\affiliation{INFN, Sezione di Trieste, I-34149 Trieste, Italy} 

\author{T. R. Zharaspayev} 
\affiliation{MEPhI: National Research Nuclear University MEPhI, RU-115409, Moscow, Russia} 

%
%
%



\begin{abstract}
We present a study of the solar-cycle variations of $>$80 MeV proton flux intensities in the lower edge of the inner radiation belt, based on the measurements of the Payload for Antimatter Matter Exploration and Light-nuclei Astrophysics (PAMELA) mission. The analyzed data sample covers an $\sim$8 year interval from 2006 July to 2014 September, thus spanning from the decaying phase of the 23rd solar cycle to the maximum of the 24th cycle. We explored the intensity temporal variations as a function of drift shell and proton energy, also providing an explicit investigation of the solar-modulation effects at different equatorial pitch angles. PAMELA observations offer new important constraints for the modeling of low-altitude particle radiation environment at the highest trapping energies.
\end{abstract}




\section{Introduction} \label{sec:Introduction}
The low-altitude ($\lesssim$1000 km) region is of special interest to the modeling of the near-terrestrial space environment, given the numerous robotic and manned missions in low-Earth orbit (LEO). The main contribution to the radiation exposure comes from the passage through the so-called South-Atlantic anomaly (SAA), where the inner Van Allen belt reaches its minimum distance from the terrestrial surface. Besides constituting a significant hazard to human activities in space, the intense fluxes of charged particles trapped in the radiation belts influence the chemical balance and energy budget of the upper atmosphere that, in turn, strongly impact on the dynamics of magnetospheric particles. Specifically, the low-altitude geomagnetically-trapped proton population is known to be coupled to the atmospheric neutral density through changes induced by the solar activity.

In general, both source and loss processes associated with inner-belt protons experience long-term variations during the solar cycle. Protons with energies in excess of a few tens of MeV mostly originate through the Cosmic-Ray Albedo Neutron Decay (CRAND) mechanism \citep{PhysRevLett.1.181,https://doi.org/10.1029/JA076i034p08223}. However, because the CRAND source for the trapped protons comes mostly from low geomagnetic latitudes (high cutoff rigidities), the solar modulation of the parent cosmic rays has a relatively small effect on their variability \citep{https://doi.org/10.1029/2006SW000275}. On the other hand, a major role is played by the changes in the atmospheric loss processes, including ionization and scattering off neutral and ionized atoms, induced by the solar extreme ultraviolet (EUV) radiation heating the Earth's upper atmosphere. In fact, the increased EUV output during solar maxima leads to higher neutral and ionospheric drift-averaged densities, with a consequent decrease of proton flux intensities. These changes depend sensitively on the drift shell, as ionospheric and atmospheric losses become less effective with increasing altitude, thus resulting in longer particle lifetimes \citep{doi:10.1063/1.1705940,https://doi.org/10.1029/JZ069i019p03927,PhysRevLett.20.806,https://doi.org/10.1029/JA076i010p02313,https://doi.org/10.1029/JA085iA01p00001,doi:https://doi.org/10.1029/GM097p0119,HUSTON19981625,https://doi.org/10.1029/1999GL003721,https://doi.org/10.1029/2006SW000275,Kuznetsov2010}. 

Our current knowledge of the solar-cycle variations of the low-altitude proton environment essentially relies on the long-term measurements below $\lesssim$250 MeV made by a few LEO missions, including the NOAA Polar Orbiting Environmental Satellite (since 1978) series at an $\sim$800--850 km altitude (e.g., \citet{doi:https://doi.org/10.1029/GM097p0119,HUSTON19981625,https://doi.org/10.1002/2014JA020300}), and the Solar, Anomalous, and Magnetospheric Particle Explorer (1992--2012), launched into a $\sim$510--690 km altitude orbit (e.g., \citet{https://doi.org/10.1029/2020JA028198}). These experiments have reported up to order-of-magnitude variations in the innermost region of the belt, while no significant modulation was observed at McIlwain's $L$ $\gtrsim$ 1.2 R$_{\mathrm E}$. The proton flux was found to rapidly decrease in the solar-maximum phases and slowly increase approaching solar minima. In particular, it was shown that the intensity variations are anticorrelated with those of the solar radio flux at 10.7 cm (also called the F$_{10.7}$ index), commonly used as a proxy of the corona activity and as an input to atmospheric models, with an energy- and $L$-dependent phase lag \citep{doi:https://doi.org/10.1029/GM097p0119,HUSTON19981625,https://doi.org/10.1029/2020JA028198}. 

In this letter we present a new study based on the data of the Payload for Antimatter Matter Exploration and Light-nuclei Astrophysics (PAMELA) experiment, which has recently offered the opportunity to
improve the description of the proton fluxes in LEO, extending it to the highest trapping energies. 

\section{The PAMELA experiment} \label{sec:The PAMELA experiment}
The PAMELA detector was conceived for a precise measurement of charged cosmic rays -- protons, electrons, their antiparticles and light nuclei -- with energies between several tens of MeV and several hundreds of GeV \citep{ADRIANI2014323}. The instrument comprised a magnetic spectrometer with a microstrip silicon tracker, a time-of-flight system shielded by an anticoincidence system, an electromagnetic calorimeter and a neutron detector. The $Resurs$-$DK1$ Russian satellite, which hosted the apparatus, was launched into a semi-polar (70 deg inclination) and elliptical (350$\times$610 km altitude) orbit in 2006 June; the perigee was subsequently raised to a nearly circular orbit of 565--590 km altitude in early 2010 September. During almost 10 years of continuous data-taking, PAMELA has been providing invaluable observations of the near-Earth radiation environment, including galactic cosmic-rays -- and their solar modulation \citep{NUOVOCIMENTO}, solar energetic particle events \citep{Bruno_2018} and magnetospheric particles \citep{BRUNO2017788}. In particular, PAMELA has enabled a comprehensive investigation of albedo and geomagnetically-trapped proton \citep{Adriani_2015,https://doi.org/10.1002/2015JA021019}, antiproton \citep{Adriani_2011}, electron and positron \citep{https://doi.org/10.1029/2009JA014660,Mikhailov2020} populations in LEO.

\section{Data Analysis} \label{sec:Data Analysis}
The analysis presented in this work is based on the proton data acquired between 2006 July and 2014 September in the SAA, defined as the geomagnetic region characterized by a field strength $B$$<$0.23 G, based on the 12th generation International Geomagnetic Reference Field (IGRF-12) model \citep{IGRF12}. No attempt was made to separate stably-trapped protons from the much less intense loss-cone (quasi-trapped and un-trapped) populations, while galactic, solar and penumbra protons were rejected by selecting only protons with magnetic rigidities $R$$<$10/$L^{3}$ GV \citep{Adriani_2015,https://doi.org/10.1002/2015JA021019}. Flux intensities were reconstructed by accounting for both pitch-angle and East-West anisotropies, particularly relevant at PAMELA altitudes and energies (see \citet{BRUNO2021} for details).

Based on statistical considerations, the solar-cycle variations were investigated by dividing the analyzed data sample into 12 temporal intervals. Specifically, to avoid orbital effects associated with the trapped-flux spatial gradient and anisotropy, the dataset acquired before the orbit change was divided into six time bins, each of $\sim$224 days, approximately corresponding to the $Resurs$-$DK1$ orbit precession period. For consistency, a similar subdivision was applied to the data collected after 2010 September in nearly circular orbit, with six intervals of $\sim$245 days. We excluded the incomplete period between late 2010 February and early 2010 September, characterized by relatively low selection efficiencies and affected by some data gaps. The time variations of the PAMELA detector performance were taken into account by estimating the selection efficiencies and related uncertainties in each temporal bin. 

We focused on $L$-shells below 1.22 R$_{\mathrm E}$, accumulating the data into three intervals: 1.10--1.13 R$_{\mathrm E}$, 1.13--1.17 R$_{\mathrm E}$ and 1.17--1.22 R$_{\mathrm E}$. Consistent with the PAMELA angular resolution, data were subdivided into 3-deg equatorial pitch-angle $\alpha_{eq}$ bins. Due to the altitude orbital constraints, the observable $\alpha_{eq}$ range varied with the $L$-shell and, in particular, PAMELA was able to detect equatorially-mirroring protons up to $\sim$1.18 R$_{\mathrm E}$. Consequently, it was not possible to measure fluxes in the 87--90 deg pitch-angle bin for the highest $L$ interval with adequate precision.

\begin{figure}[!t]
\centering
\includegraphics[width=0.65\textwidth]{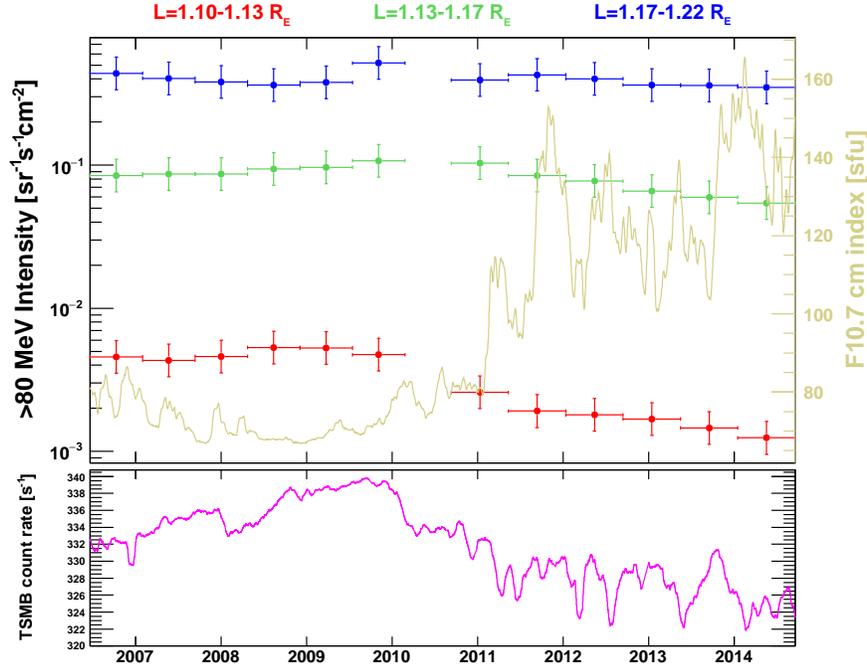} \\
\caption{Top panel: time variations of $>$80 MeV proton intensities measured by PAMELA between 2006 July and 2014 September, in three consecutive $L$-shell bins (color code); for comparison, the dark-yellow curve marks the solar 10.7 cm radio flux during the same period. Bottom panel: count rate of the Tsumeb neutron monitor, with geomagnetic-cutoff rigidity comparable to PAMELA observations. Both F$_{10.7}$ and Tsumeb data are based on monthly running averages.}
\label{fig:RadioComparison}
\end{figure}

The temporal profiles of the $>$80 MeV proton intensities, averaged over the pitch-angle, are shown in the top panel of Figure \ref{fig:RadioComparison}; the color code refers to the three aforementioned $L$ bins. Each point marks the center of the corresponding temporal interval, with the horizontal and the vertical bars denoting the bin width and the total uncertainty ($\gtrsim$30\%), respectively. For comparison, the dark-yellow curve marks the corresponding variations of the solar 10.7 cm radio flux during the same period. In addition, the bottom panel of Figure \ref{fig:RadioComparison} reports the pressure- and efficiency-corrected count rate of the Tsumeb neutron monitor station in Namibia, with a geomagnetic-cutoff rigidity $R_{c}$ $\sim$ 9.15 GV comparable to the PAMELA observations.

\begin{figure}[!t]
\centering
\includegraphics[width=\textwidth]{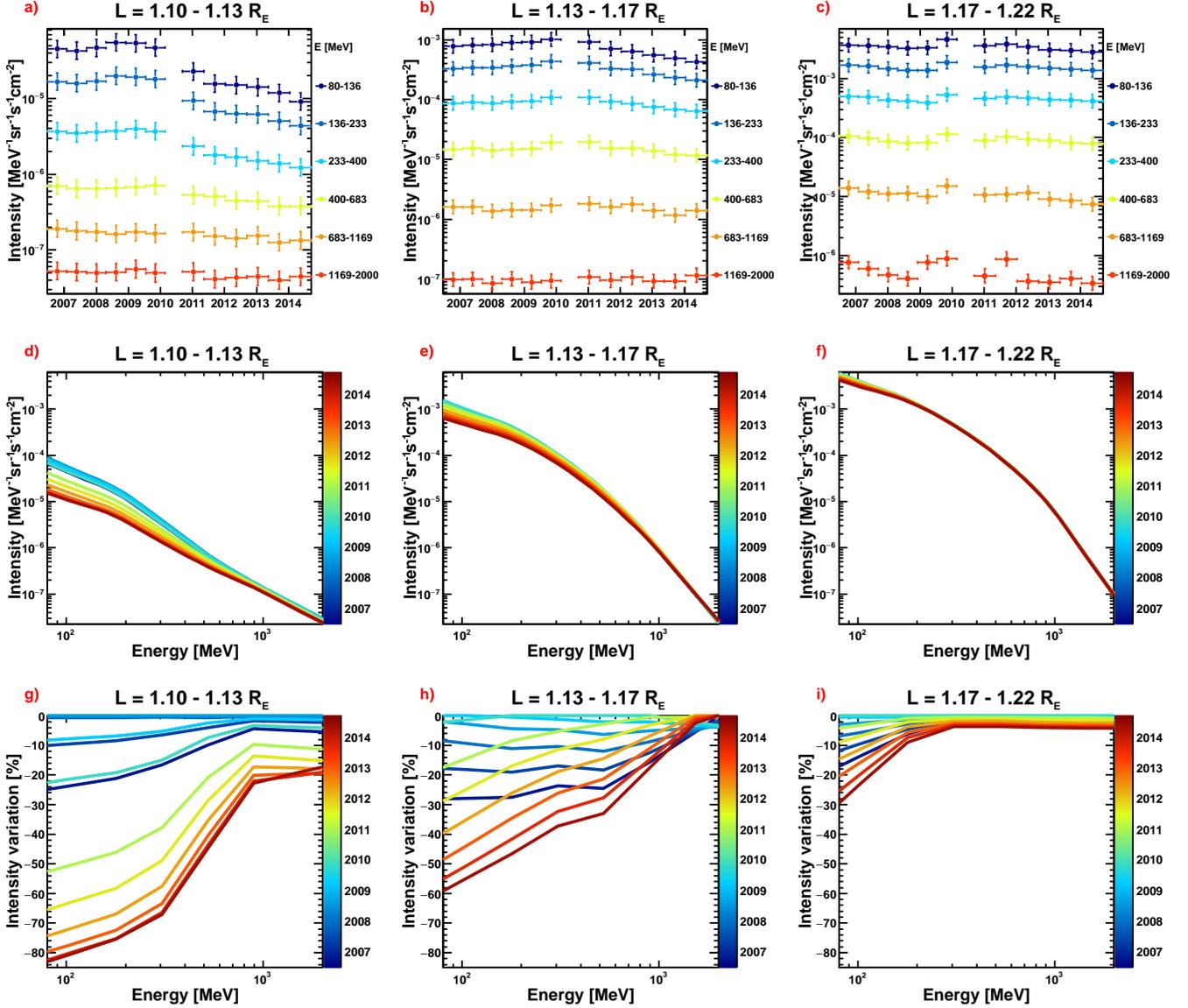} \\
\caption{Top panels: temporal profiles of pitch-angle averaged proton intensities as a function of energy (color code), for thee different $L$-shell bins.
Middle panels: resulting proton energy spectra as a function of time (color code). Bottom panels: corresponding intensity variations relative to the highest flux intensity in each energy bin. Lines are to guide the eye.}
\label{fig:Spectra}
\end{figure}

The energy dependence of flux solar-cycle variations is demonstrated in Figure \ref{fig:Spectra}. The top panels display the differential intensities measured in six energy bins between 80 MeV and 2 GeV (color code); the middle panels show the resulting energy spectra, with the color code indicating the observation time; finally, the bottom panels report the intensity relative variations $\Delta J$=$(J_{min}-J_{max})/J_{max}$, with $J_{min}$ and $J_{max}$ being the minimum and maximum intensity values during the covered time interval, respectively. We note that, for simplicity, measurement uncertainties are only shown in the top panels.

\begin{figure}[!t]
\centering
\includegraphics[width=\textwidth]{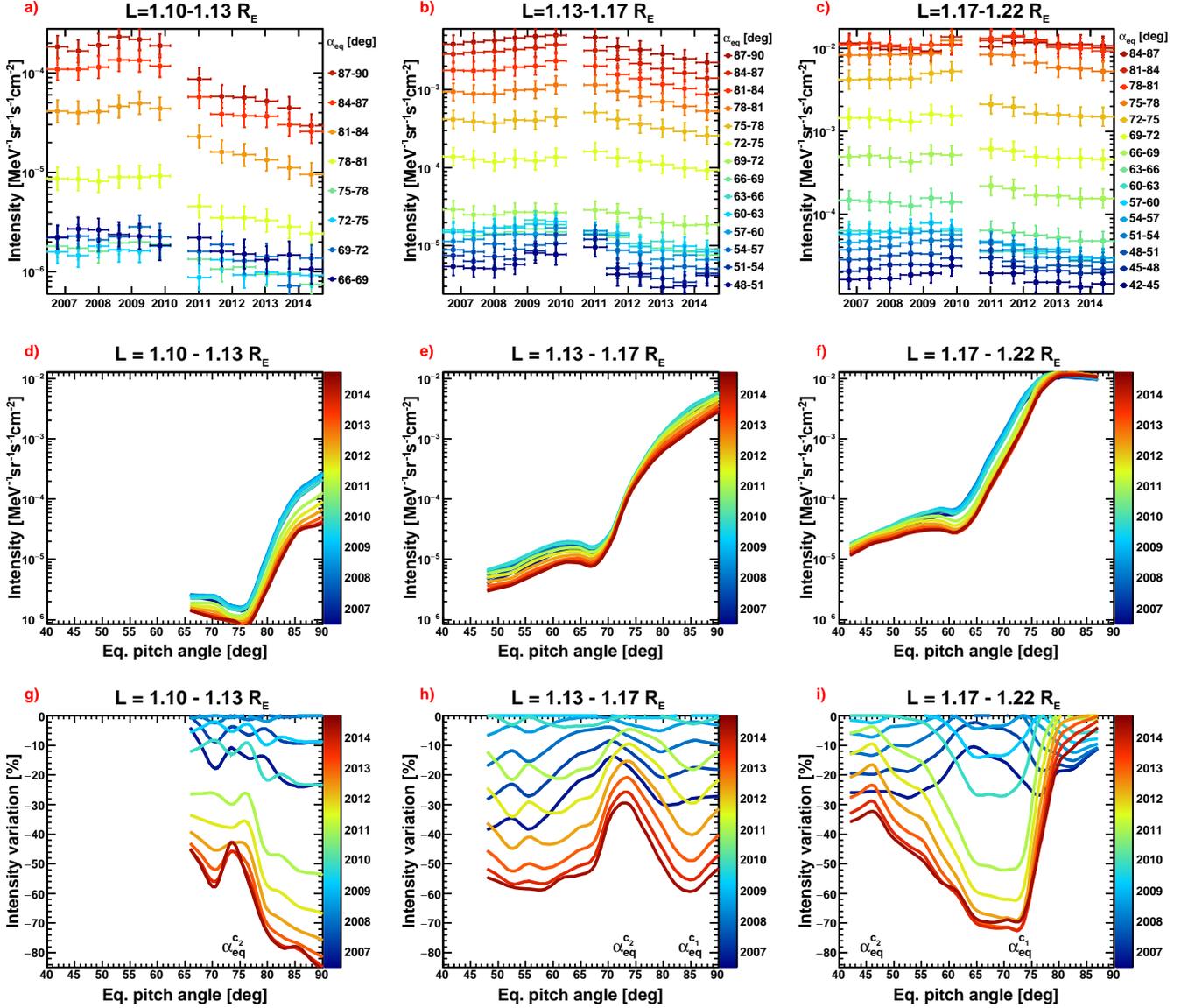} \\
\caption{Top panels: temporal profiles of 80--136 MeV proton intensities as a function of equatorial pitch angle (color code), for three different $L$-shell bins. 
Middle panels: resulting equatorial pitch angle profiles as a function of time (color code). Bottom panels: corresponding intensity variations relative to the highest intensity in each pitch-angle bin;
the approximate values of the critical pitch angles ($\alpha_{eq}^{c_{1,2}}$) are also indicated (see the text for details).
Lines are to guide the eye.}
\label{fig:pitchdistr}
\end{figure}

Finally, the solar-modulation effects on the pitch-angle distributions of 80--136 MeV protons are investigated in Figure \ref{fig:pitchdistr}.
Similar to Figure \ref{fig:Spectra}, the top panels display the temporal profiles of intensities for different pitch-angle bins (color code);
the middle panels show the resulting pitch-angle profiles, with the color-code indicating the observation time; finally, the bottom panels demonstrate the relative variations with respect to the peak values in each pitch-angle bin.

\section{Discussion and Conclusions} \label{sec:Discussion and Conclusions}
The PAMELA experiment has enabled the possibility of investigating the temporal variations of trapped proton fluxes in the kinetic interval ranging from 80 MeV to the highest trapping energies (2 GeV). 
The analyzed data sample covers the $\sim$8 year interval from 2006 July to 2014 September, thus spanning from the extended solar-minimum phase between of the 23rd and the 24th cycles, to the delayed maximum of the 24th cycle, characterized by a relatively low solar activity compared to previous solar maxima. Explored altitudes include the innermost portion of the belt, where collisions with atmospheric neutrals are dominant. 

As discussed below, reported low-$L$ intensities are characterized by a much larger variation with respect to the solar modulation of parent cosmic rays ($\lesssim$5\%), as inferred from the count-rate observations of the Tsumeb neutron monitor station at a similar geomagnetic cutoff (see Figure \ref{fig:RadioComparison}). The flux temporal profiles exhibit the well-know anti-correlation with the F$_{10.7}$ index, commonly used as a proxy for the solar EUV input to the Earth's upper atmosphere. In particular, the proton intensity drop following the radio-flux increase between late 2010 and early 2011 is evident. Unfortunately, the PAMELA data gaps in 2010, along with the relatively low temporal resolution used in this analysis, precluded a precise investigation of the trapped-flux peak times and their time lag with respect to the solar radio emission. 

As expected based on the faster energy-loss timescales, the solar-cycle variations demonstrated in Figure \ref{fig:Spectra} are larger for lower-energy protons and lower $L$ values. In particular, the variation amplitude $|\Delta J|$ at $L$=1.10--1.13 R$_{\mathrm E}$ spans from $\sim$80\% at 80--136 MeV to $<$20\% above 1 GeV (see panel ``g''), reflecting the different rates associated with ionization and free electron losses \citep{https://doi.org/10.1029/2006SW000275}. A similar variation is observed moving from the lowest to the highest $L$-shell bins, reflecting the corresponding drift-averaged lifetime increase with increasing altitude, resulting in smaller $|\Delta J|$ while approaching the solar-cycle duration \citep{https://doi.org/10.1029/1999GL003721,Kuznetsov2010}.

On the other hand, the temporal variations estimated for the pitch-angle distributions reported in Figure \ref{fig:pitchdistr} are less obvious. Since particles mirroring at lower altitudes experience larger variations associated with a larger atmospheric absorption, the variation amplitude is expected to increase with decreasing pitch angles. However, this trend is observed only in the highest pitch-angle regions. These results can be interpreted, at least in part, by accounting for the fact that measured fluxes include loss-cone (quasi-trapped and precipitating) particle components \citep{Adriani_2015,https://doi.org/10.1002/2015JA021019}.
 
Quasi-trapped protons are created and absorbed by the atmosphere within a few drift cycles (accounting for finite gyroradius effects) in the equatorial regions extending westward and eastward of the SAA, respectively (see Figure 2, top panels, in \citet{https://doi.org/10.1002/2015JA021019}), so they are not expected to experience large solar-cycle variations. Furthermore, their population will be replenished by the stably-trapped protons injected in the drift loss cone due to the increased atmospheric absorption at solar maximum. Consequently, the variation amplitude estimated for the total (stably-trapped plus quasi-trapped) flux will stop increasing with decreasing pitch angle after reaching a maximum at some critical value $\alpha_{eq}^{c_{1}}$ (see bottom panels in Figure \ref{fig:pitchdistr}). 

Precipitating protons are instead created and absorbed by the atmosphere within a bounce period, thus detected at the same longitudes. Both sink and origin point distributions are peaked in the SAA, but the latter has an additional peak in the northern hemisphere corresponding to particles with southern mirror points in the SAA (see Figure 2, middle panels, in \citet{https://doi.org/10.1002/2015JA021019}). Protons produced in the SAA with a pitch angle greater than a critical value $\alpha_{eq}^{c_{2}}$ can undergo a reflection in the northern hemisphere (where the mirror point altitude is higher) before being reabsorbed in the SAA; conversely, protons with a lower $\alpha_{eq}$ will have both mirror points at relatively low altitudes, and will be absorbed within half bounce period. The value $\alpha_{eq}^{c_{2}}$ is longitude dependent as a consequence of the longitudinal dependence of the mirror point altitude. The extended scale height of the atmosphere at solar maximum causes an increase of $\alpha_{eq}^{c_{2}}$, so that a larger fraction of protons is absorbed in the northern hemisphere before being reflected back to the SAA, resulting in a decrease of the measured flux intensities. 
 
As a result of these effects, including their dependence on the magnetic latitude, we observe a different trend of the variation amplitude depending on the pitch angle range. $|\Delta J|$ increases with decreasing pitch angle both in the highest pitch-angle ($\alpha_{eq}$$>$$\alpha_{eq}^{c_{1}}$) interval dominated by stably-trapped protons, and in the lowest pitch-angle ($\alpha_{eq}$$<$$\alpha_{eq}^{c_{2}}$) region mostly populated by precipitating protons. $|\Delta J|$ will reach a local minimum in the intermediate $\alpha_{eq}$ range. The three regions will shift to lower pitch-angle values for higher $L$-shells due to the increase in the mirror point altitude (see bottom panels in Figure \ref{fig:pitchdistr}). Specifically, $\alpha_{eq}^{c_{1}}$ decreases from $\sim$90 deg in the lowest $L$ bin (see panel ``g''), where the stably-trapped flux is relatively low, to $\sim$75 deg in the highest $L$ bin (see panel ``i'') where the inner belt peaks. $\alpha_{eq}^{c_{2}}$ is quite similar in the first two $L$-shell bins ($\sim$75 deg), but it significantly decreases to $\sim$45 deg at highest $L$. The time variations of the pitch-angle distributions will be further investigated by extending our analysis to the loss-cone populations in the whole equatorial region.

Finally, we note that secular changes occurring in the geomagnetic field cannot be neglected during the relatively long time interval covered by PAMELA observations. In fact, the current decrease in the Earth's dipole moment causes a lowering of mirror-point altitudes along with a slight increase of the trapped proton energies \citep{https://doi.org/10.1029/2006SW000275}.

\begin{acknowledgments}
The authors thank S.~L.~Huston, T.~P.~O'Brien and R.~S.~Selesnick for helpful discussions. They acknowledge support from The Italian Space Agency (ASI) under the program ``Programma PAMELA - attivit\`a scientifica di analisi dati in fase E'', Deutsches Zentrum f\"{u}r Luft- und Raumfahrt (DLR), The Swedish National Space Board, The Swedish Research Council, and The Russian Space Agency (Roscosmos). A. Bruno acknowledges NASA support through NESC TI-18-01404. M.~Martucci, F.~S.~Cafagna and R.~Sparvoli acknowledge ESA/ESTEC support through contract 4000129294/19/NL/AS. V.~V.~Malakhov and A.~G.~Mayorov acknowledge support from the Russian Science Fundation through project 19-72-10161. The neutron monitor data were provided by the Neutron Monitor DataBase (NMDB, \url{https://www.nmdb.eu/}). The F$_{10.7}$ solar index data were obtained from the OMNI database (\url{http://omniweb.gsfc.nasa.gov}).
\end{acknowledgments}


\bibliographystyle{aasjournal}

\end{document}